\documentclass[prd,showpacs,showkeys,preprintnumbers,twocolumn,a4paper,floatfix]{revtex4}

\usepackage[colorlinks=true, %
  urlcolor=blue,   
  citecolor=blue,  
  linkcolor=red,   
  bookmarksopen=true]{hyperref}

\usepackage{ucs}
\usepackage[utf8x]{inputenc}
\usepackage{amsmath,amssymb}
\usepackage{graphicx,psfrag,subfig}
\usepackage{rotating}
\usepackage{multirow}
\usepackage{dcolumn}

\usepackage{xspace}
\def\eg{{\it e.g.\ }}
\def\LHC{{\small LHC}\xspace}
\def\SUSY{{\small SUSY}\xspace}
\def\WMAP{{\small WMAP}\xspace}
\def\EWSB{{\small EWSB}\xspace}
\def\bsg{\ensuremath{b\to s \gamma}\xspace}

\def\M2{\ensuremath{M_{2}}\xspace}
\def\m0{\ensuremath{m_{0}}\xspace}
\def\SMgr{SU(3)$\times$SU(2)$\times$U(1)\xspace}
\def\SU51{SU(5)$\times$U(1)\xspace}
\def\SUPS{SU(4)$\times$SU(2)$\times$SU(2)\xspace}
\def\SUSO7{SU(2)$\times$SO(7)\xspace}
\def\hc{\textrm{\scriptsize H.C.}}

\def\vev{\emph{vev}}
\def\lsim{\mathrel{\rlap{\raise 2.5pt \hbox{$<$}}\lower 2.5pt\hbox{$\sim$}}}
\def\gsim{\mathrel{\rlap{\raise 2.5pt \hbox{$>$}}\lower 2.5pt\hbox{$\sim$}}}

\def\GUT{\ensuremath{{\rm G}}}

\newcommand{\refeq}[1]{Eq.~\eqref{#1}}
\newcommand{\Neu}[1]{\ensuremath{\widetilde \chi_{#1}^0}}
\newcommand{\Cha}[1]{\ensuremath{\widetilde \chi_{#1}^\pm}}
\newcommand{\wbar}[1]{\mkern1mu\overline{\mkern-3mu#1\mkern-1mu}\mkern2mu}
 
\psfrag{m0}{$m_{0}$}
\psfrag{m2}{$M_{2}$}
\psfrag{mA}{${A_0}$}
\psfrag{to}{$\tilde t_1$} 
\psfrag{ta}{$\tilde \tau_1$} 
\psfrag{my}{$\tilde \mu_1$} 
\psfrag{ch}{$\Cha 1$} 
\psfrag{n2}{$\Neu 2$} 
\psfrag{n1}{$\Neu 1$} 
\psfrag{ew}{$\textsf{rge}$} 
\psfrag{bi}{\footnotesize$\widetilde B$}
\psfrag{wi}{\footnotesize$\widetilde W$}
\psfrag{hi}{\footnotesize$\widetilde H$}

\begin{document}

\preprint{HIP-2008-41/TH}
\preprint{DO-TH-08/09}

\title{Relic density in nonuniversal gaugino mass models with SO(10)
  GUT symmetry}

\author{Katri Huitu}
 \email{katri.huitu@helsinki.fi}
 \affiliation{Department of Physics, and 
    Helsinki Institute of Physics, \\P.O.~Box 64,
    {FIN-00014} University of Helsinki, Finland}
\author{Jari Laamanen}
 \email{jari.laamanen@uni-dortmund.de}
 \affiliation{Institut f\"ur Physik, 
    Technische Universit\"at Dortmund, {D-44221} Dortmund, Germany}

\date{\today}


\begin{abstract}
  Nonuniversal boundary conditions in grand unified theories can lead
  to nonuniversal gaugino masses at the unification scale.  In
  $R$-parity preserving theories the lightest supersymmetric particle
  is a natural candidate for the dark matter. The composition of the
  lightest neutralino and the identity of the next-to-lightest
  supersymmetric particle are studied,
  when nonuniversal gaugino masses come from representations of
  SO(10).  In these cases, the thermal relic density compatible with
  the Wilkinson Microwave Anisotropy Probe observations is found.  Relic densities are compared with the
  universal case.  Mass spectra in the studied cases are discussed.
\end{abstract}


\maketitle


\section{Introduction}
\label{intro}

The phenomenology of supersymmetric models depends crucially on the
compositions of neutralinos and charginos, if the lightest neutralino
is the lightest supersymmetric particle (LSP).  In addition to the
laboratory studies, relevant input is obtained from the dark matter
searches, where the the Wilkinson Microwave Anisotropy Probe (\WMAP) satellite has put precise limits on the
relic density. Supersymmetric theories which preserve $R$-parity
contain a natural candidate for the cold dark matter particle.
Neutralino LSP can provide the appropriate relic density.

In many supergravity type models the lightest neutralino is binolike,
which often leads to too large thermal relic density, as compared to
the limits provided by the \WMAP\ experiment
\cite{Spergel:2006hy}. When the gaugino masses are not universal at
the grand unification scale, the resulting neutralino composition
changes from the case of universal gaugino masses \cite{Huitu:2005wh}.
In this paper, the thermal relic density of the neutralino LSP is
studied, when gaugino masses are due to nonuniversal representations
of SO(10) grand unified theory (GUT)
\cite{Fritzsch:1974nn,Zhao:1981me}.  Dark matter in a particular gauge
symmetry breaking chain of the SO(10) GUT in the case of universal
gaugino masses has been recently studied in \cite{Drees:2008tc}.  Some
phenomenological aspects of SO(10) GUTs with nonuniversal gaugino
masses have been considered in \cite{Bhattacharya:2007dr,Ananthanarayan:2007fj}.

SO(10) has many attractive features among the GUT models.  One of the
most appealing properties is that one family of matter fermions can be
put into a single 16-dimensional irreducible spinor representation of
SO(10), including the right-handed neutrino
\cite{Mohapatra:1979ia,Bajc:2001fe}. In addition, SO(10) allows
possibility for the Yukawa coupling unification and representations
are anomaly free.  Conservation of $R$-parity, which forbids the
unwanted dimension-five operators leading to rapid proton decay, may
result from the SO(10) symmetry breaking.  The doublet-triplet
splitting could be achieved using, \eg, the so-called
Dimopoulos-Wilczek mechanism \cite{DimopoulosWilczek}.  Because the
SO(10) gauge symmetry breaks down to the standard model (SM) gauge
symmetry through some intermediate group, the SO(10) GUT offers
several possibilities for the model building.  For example, it can
contain as a subgroup the Pati-Salam \SUPS model.


Gaugino masses originate from the non-renormalizable terms in the
$N=1$ supergravity Lagrangian involving the gauge kinetic function
$f_{ab}(\Phi)$ \cite{Cremmer:1982wb}. The gauge part of the Lagrangian
contains the gauge kinetic function coupling with two field strength
superfields $W^a$. The Lagrangian for the coupling
can be written as
\begin{eqnarray}
  {\cal L}_{gk} = \int d^2 \theta f_{ab}(\Phi) W^a W^b + \hc,
  \label{gauge-fs-kinetic}
\end{eqnarray} 
where $a$ and $b$ are gauge group indices (for example, $a,b =
1,2,...,45$ for ${\rm SO(10)}$), and repeated indices are summed over.
The function $f_{ab}(\Phi)$ is an analytic function of the chiral
superfields $\Phi$ in the theory.  The chiral superfields $\Phi$
consist of a set of gauge singlet superfields $\Phi^s$ and gauge
nonsinglet superfields $\Phi^n$ under the grand unified group.
The gauge kinetic function $f_{ab}(\Phi)$ can
be expanded,
\begin{eqnarray}
  f_{ab}(\Phi) = f_0(\Phi^s)\delta_{ab} + \sum_n f_n(\Phi^s)
  \frac{\Phi^n_{ab}}{M_P} + \cdots,
  \label{gauge-kinetic}
\end{eqnarray}
where $\Phi^s$ and $\Phi^n$ are the singlet and nonsinglet chiral
superfields, respectively. Here $f_0(\Phi^s)$ and $f_n(\Phi^s)$ are
functions of gauge singlet superfields $\Phi^s$, and $M_P$ is some
large scale.
In order to generate a mass term for the gauginos, the gauge kinetic
function must be non-minimal, i.e., it must not be a constant
\cite{Ferrara:1982qs}.
When $F_\Phi$ gets a vacuum expectation value (\vev) $\langle F_\Phi
\rangle$, the interaction (\ref{gauge-fs-kinetic}) gives rise to
gaugino masses:
\begin{eqnarray}
  {\cal L}_{gk} \supset \frac{{\langle F_\Phi \rangle}_{ab}}
  {M_P}\lambda^a \lambda^b + \hc,
  \label{gmassterm}
\end{eqnarray}
where $\lambda^{a,b}$ are gaugino fields.  The nonuniversal gaugino
masses are generated by the nonsinglet chiral superfield $\Phi^n$ that
appears linearly in the gauge kinetic function $f_{ab}(\Phi)$ in
Eq.~\eqref{gauge-kinetic}.


Gauginos belong to the adjoint representation of the gauge group,
which in the case of ${\rm SO(10)}$ is the {\bf 45} dimensional
representation.  Because \refeq{gmassterm} must be gauge invariant,
$\Phi$ and $F_\Phi$ must belong to some of the following
representations appearing in the symmetric product of the two {\bf 45}
dimensional representations of ${\rm SO(10)}$
\cite{Slansky:1981yr,Chamoun:2001in}:
\begin{eqnarray}
  ({\bf 45 \otimes 45})_{Symm} = 
  {\bf 1 \oplus 54 \oplus 210 \oplus 770}.
   \label{symmprod}
\end{eqnarray}
The representations {\bf 54}, {\bf 210} and {\bf 770} may lead to
nonuniversal gaugino masses, while the {\bf 1} dimensional
representation gives manifestly the universal gaugino masses.  The
relations between the gaugino masses are determined by the
representation invariants, and are specific for each of the
representations. Because the gauge kinetic function in
Eq.~(\ref{gauge-kinetic}) can get contributions from several different
$\Phi$'s, a linear combination of any of the representations is also
possible. In that case the gaugino mass terms are not  uniquely
determined anymore, in contrast to the contribution purely from one
representation.  Here we assume that the dominant component of the
gaugino masses comes from only one representation.  This gives us
a clear understanding of the role of different representations.

\section{Dark matter in SO(10) representations}
\label{sec:DM}
 
\subsection{Breaking Chains: SO(10) \texorpdfstring{$\to \boldsymbol{H}\to$}{-> H ->} SM}

The GUT group SO(10) breaks down to the standard model gauge
group \SMgr via some intermediate gauge group $H$. Therefore the
gaugino mass relations depend also on the gauge group breaking chain,
in addition to the representation invariants coming from the gauge
kinetic function.  Moreover, the intermediate breaking scale affects
also the generated gaugino masses via heavy gauge supermultiplets that
correspond to the broken generators. However, if the gauge breaking
from the GUT group to the SM group takes place at the GUT scale, these
loop-induced messenger contributions \cite{Giudice:1997ni} can be
neglected in comparison to the tree-level contributions.
Some fits to the experimental data in SO(10) GUT indicate that the two
breaking scales are very close to each other, see
\cite{Aulakh:2008sn,Aulakh:2005mw}, although realistic models exist
also with large splitting of the scales \cite{Bajc:2008dc}.  In this
work we assume that the breaking from SO(10) to the SM gauge group
happens at the GUT scale, and that the GUT breaking does not affect
the gauge coupling unification.

We will study the representations {\bf 54} and {\bf 210} in the
right-hand side of Eq.~(\ref{symmprod}).  The interesting breaking chains of
{\bf 54} and {\bf 210} are included also in the breaking chains of
{\bf 770}.  Table \ref{tab:chains} shows possible SO(10) breaking
chains \cite{Aulakh:1982sw,Chamoun:2001in,Fukuyama:2004ps}, which
include the standard model gauge group, for the two chosen
representations.  Some of the subgroups lead to universal gaugino
masses, or to massless gauginos \cite{Chamoun:2001in}, and we do not
consider them.  We will limit ourselves to the intermediate gauge
groups \SUPS, \SUSO7 and \SU51.
\begin{table}[htb]
  \caption{\label{tab:chains} Breaking chains of SO(10) 
representations {\bf 54} and {\bf 210} which include the SM gauge group.}
  \begin{ruledtabular}  
    \begin{tabular}{ccc}
    $F_\Phi$ & $H$ & Subgroup description  \\
    \hline
    \multirow{3}{*}{\bf 54} &
    $\scriptstyle{\text{SU}(4)\times \text{SU}(2)
      \times \text{SU}(2)}$ & {\small Pati-Salam}\\
    & $\scriptstyle{ \text{SU}(2)\times \text{SO}(7)}$ & \\
    & $\scriptstyle{\text{SO(9)}}$ & {\small Universal gauginos}\\
    \hline
    \multirow{4}{*}{\bf 210} &
    $\scriptstyle{\text{SU}(4)\times \text{SU}(2)\times \text{SU}(2)}$ & 
    {\small Massless gluino}\\
    & $\scriptstyle{\text{SU}(3)\times \text{SU}(2)
      \times \text{SU}(2) \times \text{U}(1)}$ & 
    {\small Massless }$\scriptstyle{\text{SU}(2)_\text{L}}${ \small gauginos}\\
    & $\scriptstyle{\text{SU}(3)\times \text{SU}(2)
      \times \text{U}(1) \times \text{U}(1)}$ & \\
    & $\scriptstyle{ \text{SU}(5)\times \text{U}(1)}$ & {\small
      ``Flipped'' }$\scriptstyle{\text{SU}(5)}$\\
  \end{tabular}
\end{ruledtabular}
\end{table}

Table \ref{tab:gaug} displays the ratios of resulting gaugino masses
at the tree level as they arise when $F_\Phi$ belongs to the
above-mentioned representations of ${\rm SO(10)}$ or singlet 
\cite{Chamoun:2001in}. The resulting 1-loop relations at the
electroweak scale are also displayed.
\begin{table}[htb]
  \caption{\label{tab:gaug} Ratios of the gaugino masses at the GUT
    scale in the normalization ${M_3}(GUT)$ = 1, and at the electroweak
    scale in the normalization ${M_3}(EW)$ = 1.} 
  \centering
  \begin{ruledtabular}  
  \begin{tabular}{cccccddc}
    $F_\Phi$ & $H$ & $M_1^\GUT$ & $M_2^\GUT$ & $M_3^\GUT$ & 
    \multicolumn{1}{c}{$M_1^{EW}$} &  \multicolumn{1}{c}{$M_2^{EW}$} & $M_3^{EW}$
    \\ \hline 
    {\bf 1} & & 1 & 1 & 1 &  0.14 & 0.29 & 1 \\
    {\bf 54} &$\scriptstyle{\text{SU}(4)\times \text{SU}(2)\times \text{SU}(2)}$ 
    & -1\hspace{3pt} & -1.5 & 1 &  -0.15 & -0.44 & 1 \\
    {\bf 54} &$\scriptstyle{\text{SU}(2)\times \text{SO}(7)}$ 
    & 1 & -7/3 & 1 &  0.15 & -0.68 & 1 \\
    {\bf 210} &$\scriptstyle{\text{SU}(5)\times \text{U}(1)}$ 
    & -96/25 & 1 & 1 &-0.56 &  0.29 & 1 \\
  \end{tabular}
\end{ruledtabular}
\end{table}
These values and the resulting relic densities can be compared with
the universal and nonuniversal representations resulting in the SU(5) GUT
model \cite{King:2007vh,Huitu:2008sa}.  Since we assume breaking at
one scale, the universal model with which we should compare in the
SO(10) GUT is similar to the universal model in the SU(5) GUT.  In
the nonuniversal representations, the relations between gaugino
masses change.  Thus, \eg, the {\bf 54}-dimensional Pati-Salam
model of SO(10) may seem at first glance rather similar to {\bf 24} of
SU(5), but we will see that the twice as large bino component has a
large effect to the relic density.  The bino and wino mass parameters
affect directly the lightest neutralino mass and properties. They also
affect the value of the $\mu$-parameter through the renormalization
group equations (RGE) and the radiative electroweak symmetry breaking
(rEWSB), therefore controlling also the Higgsino component in the
lightest neutralino.  Since the lightest neutralino mass limit can be
deduced from the chargino mass limit, the nonuniversal gaugino masses
change the lower limit for the neutralino mass: for {\bf 54} the
neutralino mass limit is smaller than in the universal case, while for
{\bf 210} the mass limit is close to the chargino mass limit.

\subsection{Calculation of Dark Matter Relic Density}

We calculate the \SUSY spectrum for each model with the program
{\textsf SOFTSUSY} (version 2.0.11) \cite{Allanach:2001kg}, and the
resulting relic density with the program {\textsf micrOMEGAs} (version
2.0.7) \cite{Belanger:2001fz,Belanger:2004yn,Belanger:2006is}.  For
the relic density, we use here the \WMAP\ combined 3 yr limits
\cite{Spergel:2006hy}
\begin{eqnarray}
  \Omega_{CDM} h^2 = 0.11054^{+0.00976}_{-0.00956} \quad (2\sigma).
\end{eqnarray}
In all the figures that we show below, the filling denoted by
\textsf{wmap} is the \WMAP-preferred region.
For the $b\to s\gamma$ experimental branching fraction, 
we have used the two sigma world average 
\cite{Barberio:2007cr},
\begin{eqnarray}
  BR(b\to s \gamma) = (355 \pm 24^{+9}_{-10} \pm 3) \times 10^{-6}.
\end{eqnarray}
The areas enclosed by the \textsf{bsg} contour are disallowed by the $b\to
s \gamma$ constraint.  For the particle masses, the following limits
are applied \cite{Belanger:2006is}: $m_{\tilde e_R} > 99.4$ or 100.5
GeV depending on if the lightest neutralino mass is below or above 40
GeV, $m_{\tilde \mu_R} > 95$ GeV, $m_{\tilde\tau_1} > 80.5$ to 88 GeV
depending on the lightest neutralino mass (from 10 to 75 GeV),
$m_{\tilde \nu_i} > 43$ GeV, and $m_{\tilde\chi^\pm} > 73.1$ to 103
GeV depending on the sneutrino masses (from 45 to 425 GeV).  In the
figures, \textsf{lep} shows an area where the experimental mass
limits are not met, \textsf{rge} shows an area where there is no
radiative \EWSB, and \textsf{lsp} the area where neutralino is not the
LSP.  The curve $m_h=114$ GeV is depicted in the figures (dash-dotted
line denoted by \textsf{h}).  For the shown parameter regions, when
otherwise experimentally allowed, Higgs is always heavier than $91$
GeV, which is the Higgs mass limit in MSSM for $\tan\beta \ge 10$
assuming maximal top mixing \cite{unknown:2001xy}.

\subsubsection{Representation 54}
\begin{figure*}
  \subfloat[Relic density]{
    \includegraphics[width=0.32\textwidth]{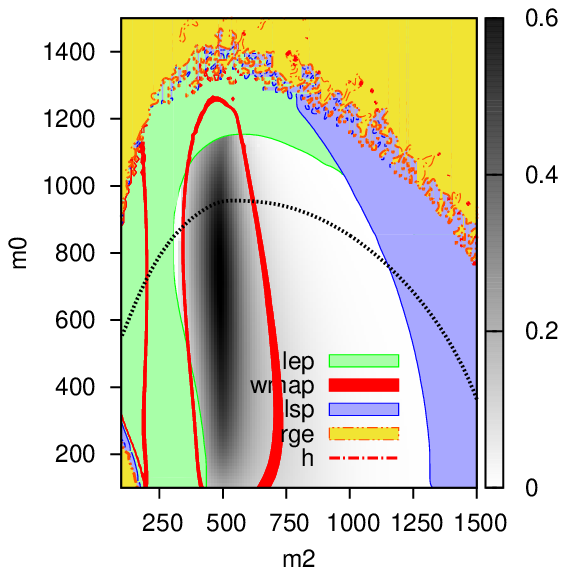}
    \label{fig:relic542m0-a}
  }  
  \subfloat[NLSP map]{
    \includegraphics[width=0.32\textwidth]{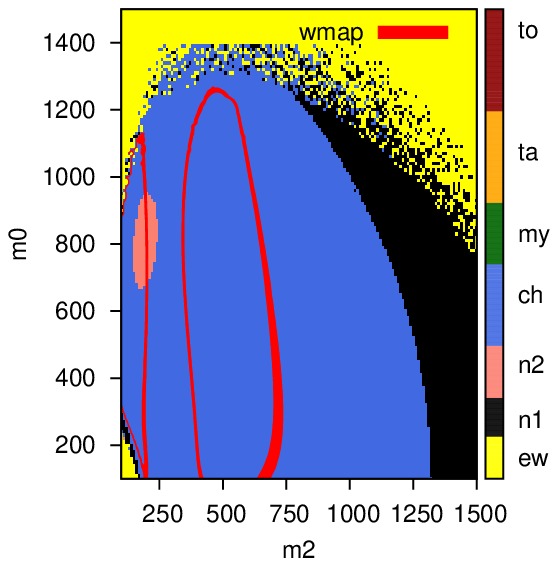}
    \label{fig:relic542m0-b}
  } 
  \subfloat[Neutralino composition]{
    \includegraphics[width=0.3\textwidth,height=60mm]{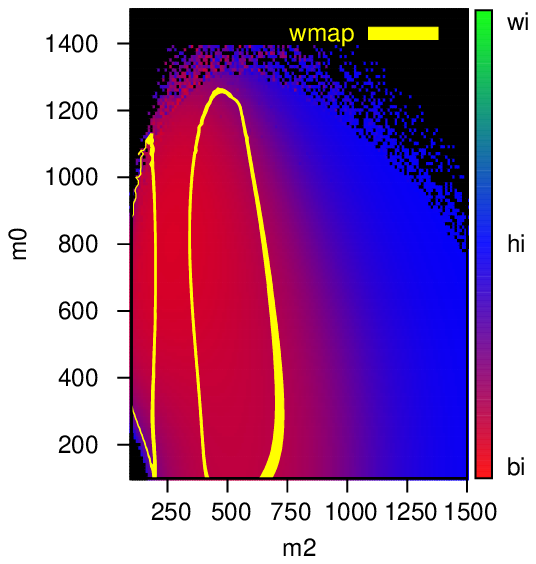}
    \label{fig:relic542m0-c}
  }
  \caption{Relic density $\Omega_\chi h^2$ in the representation
    \textbf{54} with $H=$\SUSO7 in the ($M_{2}, m_0$) plane for $\tan\beta
    = 10,\ \mathrm{sgn} ( \mu ) = +1,\ A_0=0$. In
    \ref{fig:relic542m0-a} the dark shaded areas represent the larger
    relic density.  The filling denoted by \textsf{wmap} is the \WMAP
    preferred region, \textsf{lep} shows an area, where the
    experimental mass limits are not met, \textsf{rge} shows an area
    where there is no radiative \EWSB, and \textsf{lsp} the area where
    neutralino is not the LSP.  \textsf{h} encloses the area with $m_h
    < 114$ GeV, and in the following figures \textsf{bsg} the area
    disallowed by $b \to s \gamma$ limits. In \ref{fig:relic542m0-b}
    the NLSP within the same region is plotted, and in
    \ref{fig:relic542m0-c} the $\Neu 1$ composition.  }
  \label{fig:relic542m0}
\end{figure*}

\begin{figure*}
  \subfloat[Relic density]{
    \includegraphics[width=0.32\textwidth]{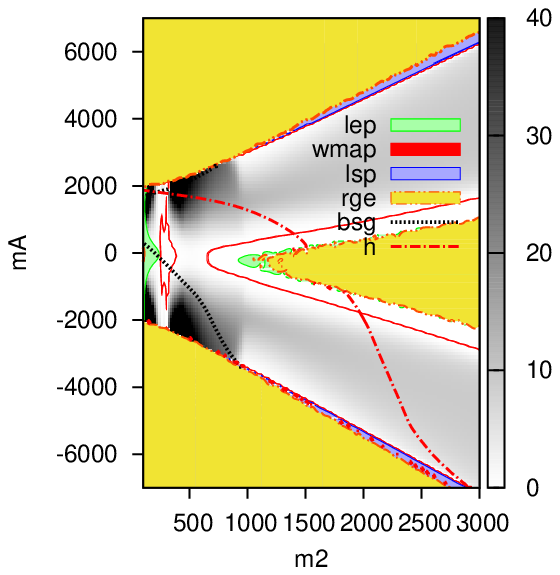}
    \label{fig:relic542A0-a}
  }  
  \subfloat[NLSP map]{
    \includegraphics[width=0.32\textwidth]{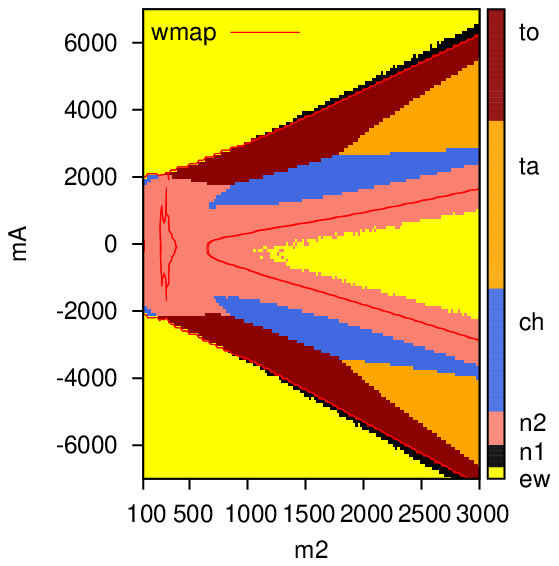}
    \label{fig:relic542A0-b}
  }
  \subfloat[Neutralino composition]{
    \includegraphics[width=0.3\textwidth,height=60mm]{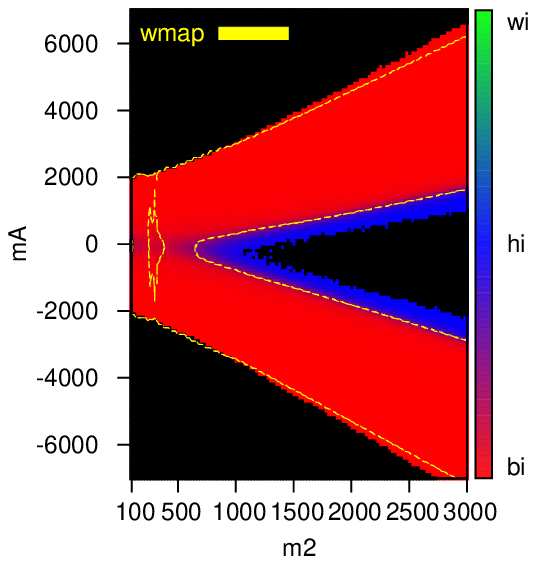}
    \label{fig:relic542A0-c}
  }
  \caption{Representation \textbf{54}: $H=$\SUSO7, $\tan\beta = 10,\
    \mathrm{sgn} ( \mu ) = -1,\ m_0=1$ TeV. Otherwise as in
    Fig. \ref{fig:relic542m0}.  }
  \label{fig:relic542-A0}
\end{figure*}

\begin{figure*}
  \subfloat[Relic density]{
    \includegraphics[width=0.32\textwidth]{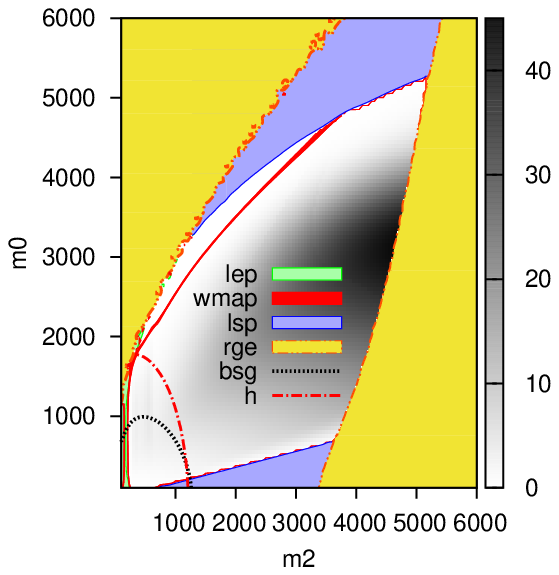}
    \label{fig:relic541m0-a} 
  }
  \subfloat[NLSP map]{
    \includegraphics[width=0.32\textwidth]{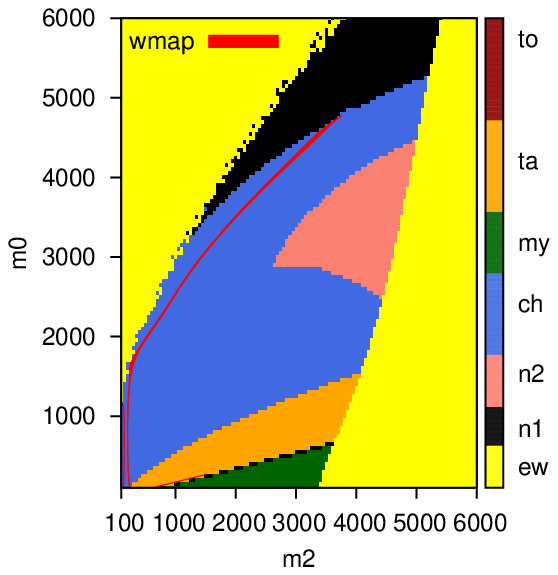}
    \label{fig:relic541m0-b}
  }
  \subfloat[Neutralino composition]{
    \includegraphics[width=0.3\textwidth,height=58mm]{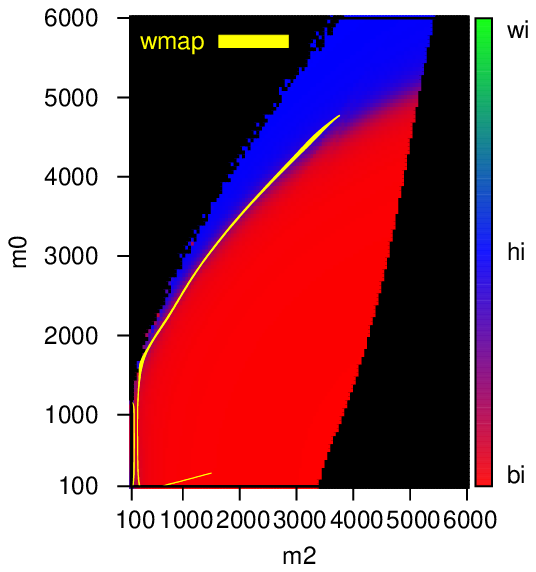}
    \label{fig:relic541m0-c}
  }
  \caption{\label{fig:relic541m0}Representation \textbf{54} with
    $H=$\SUPS in the ($M_{2}, m_0$) plane for $\tan\beta = 10,\
    \mathrm{sgn} ( \mu ) = +1,\ A_0=0$.  Otherwise as in
    Fig. \ref{fig:relic542m0}.  }
\end{figure*}

The area of preferred thermal relic density for the two chains of the
{\bf 54} dimensional representation are shown for sets of parameters
in Figs. \ref{fig:relic542m0}, \ref{fig:relic542-A0}, and
\ref{fig:relic541m0}. In each set of three figures, the first figure
(a) represents the neutralino relic density for given parameters with
collider constraints depicted in the plot, the second figure (b)
shows, for the same parameters, the identity of the next-to-lightest
supersymmetric particle, and the third figure (c) shows the lightest
neutralino composition in RGB-color encoding, ({\it i.e.}, colors, or
hues of black and white, indicate the particle as shown in the figure;
therefore the mixture of the colors, or hues of black and white,
describes the nature of the $\Neu 1$-composition).  In each of the
figures, the \WMAP-preferred relic density filling is also
superimposed to the graph.

As can be seen from the Table \ref{tab:gaug}, the lightest neutralino
is expected to be bino rather than wino. The large bino component tends to
suppress the neutralino annihilation cross section, since bino lacks
the s-channel $Z$-boson annihilation mode. A substantial Higgsino component
is usually needed to help to increase the annihilation rate, unless
there happens to be coannihilation or, \eg, an open Higgs s-channel
annihilation mode available.

In Fig.~\ref{fig:relic542m0} relic density, the next-to-lightest
supersymmetric particle (NLSP) and LSP composition in the breaking
chain \SUSO7 are shown.  Because $\Neu 1$ is mostly bino, the spectrum
with preferred relic density is quite light and conflicts with
collider constraints in some parts of the parameter space. With
increasing gaugino masses also the Higgsino component in the
neutralino LSP increases, and at the point where the change to
dominantly Higgsino LSP occurs, also the relic density drops. The
overall relic density is not very high, thus allowing a wider
\WMAP-preferred region than, \eg, in the singlet, \emph{i.e.}, mSUGRA
case \cite{Huitu:2008sa}. For a given $M_2$, the corresponding $M_1$ is
smaller than in the singlet case, which results in a smaller $\mu$
value at the EW-scale. This has an effect of an increasing the Higgsino
component in the lightest neutralino thus boosting the annihilation.
The NLSP is chargino, and with increasing Higgsino component it
eventually becomes the LSP.

In the allowed region, \M2 is less than 740 GeV, which restricts the
lightest chargino mass to values less than $\sim$ 150~GeV.  The lower
limit for the chargino mass is the LEP limit. The partners of the SM
fermions are heavier than 300--500 GeV.  Thus, assuming that the
neutralino is responsible for the dark matter, this breaking chain of
the gauge symmetry has as a robust prediction for the upper limit of
the chargino mass, and furthermore it is lighter than the squarks
and sleptons.
For \M2 $\sim$ 350~GeV the whole neutralino and chargino
spectrum, and even the gluino, is lighter than the sfermions for the LEP
allowed and \WMAP-preferred region. The same is true also for \M2
$\sim$ 630~GeV and $\m0 \gsim 940$ GeV for the \WMAP-preferred region.

The $b\to s \gamma$ constraint cuts away a considerable area from the
otherwise allowed region. Including a 10\% error in theoretical
calculations of the decay, the constraints from $b\to s \gamma$ loosen
considerably, and all of the otherwise allowed \WMAP area becomes
available. The lightest Higgs is always lighter than 114 GeV,
but heavier than 91 GeV.

The effect of varying the universal trilinear coupling $A_0$ is shown
in Fig.~\ref{fig:relic542-A0}. In contrast to the previous figure,
here the sign of the $\mu$-parameter is chosen to be negative. In
general, giving the $A_0$-parameter a nonzero value tends to increase
the relic density. Increasing $|A_0|$ will help the $m_{H_u}^2$ to run
larger negative values during the RG-evolution, and therefore to
increase the actual (absolute) value for the $\mu$-parameter via the
rEWSB. This in turn favors the bino component in the $\Neu 1$
composition in the \SUSO7 chain. The effect can be seen in the
LSP-composition Fig.~\ref{fig:relic542A0-c}.  The preferred \WMAP
region still follows the transition zone of the \Neu 1 from bino to
Higgsino.
The effect of negative $\mu$ is most visible in the fact that the NSLP
in this case is \Neu 2, since negative $\mu$ tends to increase the
lightest chargino mass (at least in the limit of $|M_2| < |\mu|$).
The \bsg constrains an area with negative $A_0$ values at $M_2 < 1$
TeV, and again including also the theoretical error, the constrained
area becomes considerably smaller.
The light Higgs boson mass is heavier than 114 GeV above the
dash-dotted line.

In Fig.~\ref{fig:relic541m0} the relic density, NLSP and \Neu 1
composition are plotted for the \SUPS breaking chain.  In this breaking
chain the $|M_1|$ and $|M_2|$ values are closer to each other at the
EW-scale than in the \SUSO7 chain, but still wider spread than in the
singlet. That has an effect of increasing the $\mu$ value, therefore
resulting in a smaller Higgsino component for the equal $M_2$ values for
the two chains of the {\bf 54} dimensional representation.  Again the
\WMAP allowed narrow region follows the transition from the \Neu 1
from bino to Higgsino.
The \bsg and Higgs boson 114 GeV limits cut pieces from near the low
$M_2, m_0$-values. The NLSP is mostly the lightest chargino, but an
interesting region exists with small $m_0$ near the area, where \Neu 1
is no longer the LSP; along the line of transition from the stau
NLSP to the smuon NLSP there is a narrow region, where \Neu 1, stau
and smuon masses are very close to each other, and the
coannihilations may reduce the relic density to an acceptable level. For
example, for $\M2=1400$ GeV, the LSP mass is around 410 GeV, and for
$\M2 = 1000$ GeV, the LSP mass is 290 GeV for the \WMAP-preferred
area.  However, this area is highly prone to numerical subtleties, and
the ordering of the LSP identity changes in the preferred relic
density range, when comparing the output of different spectrum
calculators.  Reducing the $m_0$ parameter further makes the stau
become the LSP.  From the collider point of view, such regions may be
especially interesting, as they would lead to quasistable smuons or
staus.

\subsubsection{Representation 210}
\label{subsec:rep210}
\begin{figure*}
  \subfloat[Relic density]{
    \includegraphics[width=0.32\textwidth]{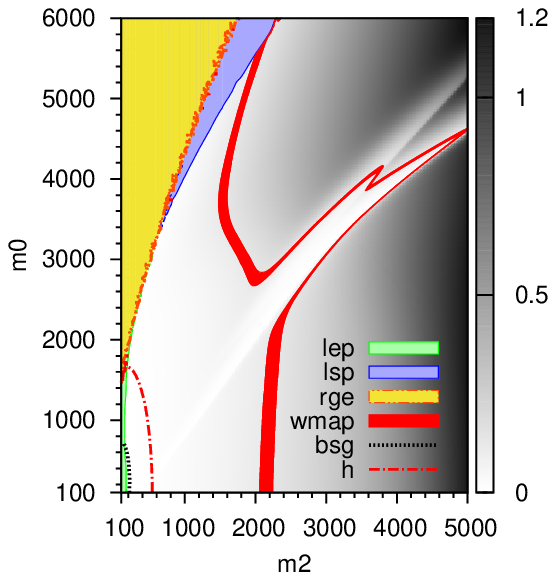}
    \label{fig:relic210-a}
  }  
  \subfloat[NLSP map]{
    \includegraphics[width=0.32\textwidth]{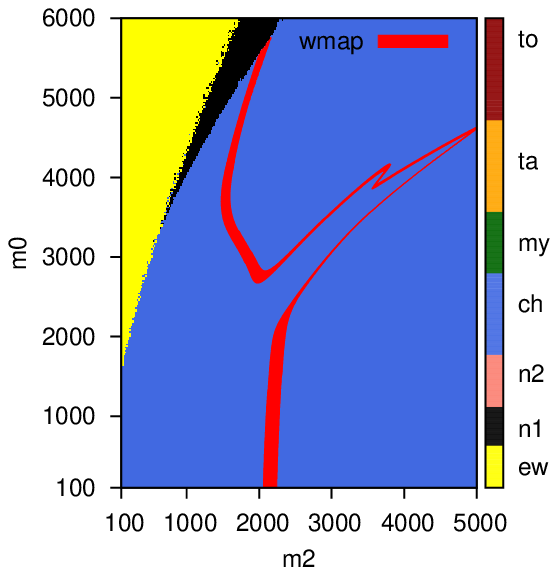}
    \label{fig:relic210-b}
  }
  \subfloat[Neutralino composition]{
    \includegraphics[width=0.3\textwidth,height=60mm]{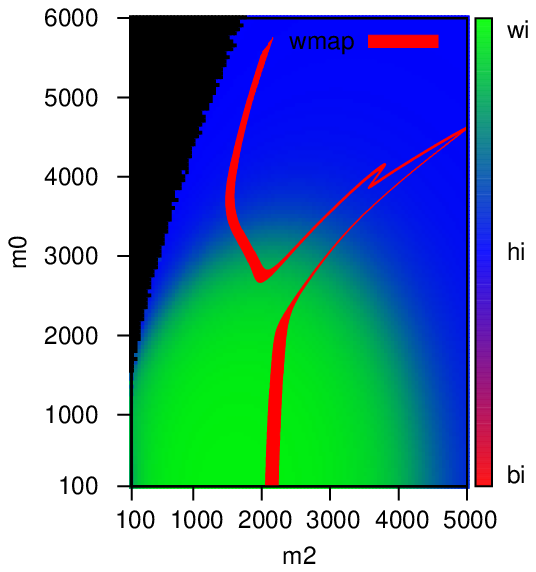}
    \label{fig:relic210-c}
  }
  \caption{Representation \textbf{210}: $\tan\beta = 10,\ \mathrm{sgn} ( \mu ) =
    +1,\ A_0=0$. Otherwise as in Fig. \ref{fig:relic542m0}.}
  \label{fig:relic210m0}
\end{figure*}

In the representation \textbf{210} we inspected the breaking chain
through the intermediate gauge group \SU51, called flipped SU(5)
\cite{DeRujula:1980qc,Barr:1981qv,Derendinger:1983aj,Antoniadis:1987dx}.
In Fig.~\ref{fig:relic210-a} the area of preferred thermal relic density in
the representation \textbf{210} is plotted for a set of (GUT scale)
parameters. For the chosen parameters, rather large \WMAP-preferred
regions are found for large values of \M2 and/or \m0 parameters. When
not Higgsino, the lightest neutralino is expected to be wino, rather
than bino (see Table \ref{tab:gaug}), and therefore the neutralino
relic density to be very small.  In general, due to the wino being
the smallest of the two electroweak gaugino parameters, it characterizes
the lightest neutralino.  Since  the lightest chargino is
characterized also by this parameter, for a large part of the parameter
space, the masses of the \Neu 1 and \Cha 1 are very close to each
other, which boosts the rapid neutralino relic density
annihilation. The situation resembles the one arising in the anomaly
mediated \SUSY breaking scenario, where also both the lightest
neutralino and chargino are characterized by the wino mass parameter.

Since both the wino and Higgsino have a large annihilation cross section,
the \WMAP-preferred relic density region does not have to follow the
transition zone of \Neu 1 from wino to Higgsino. The increase of the
relic density to the observed level is mainly due to the increase of
the mass parameters $M_2$ and $m_0$. An interesting change in the
pattern can be seen on the diagonal of the figures, where the \Neu 1
and \Cha 1, and also at some point \Neu 2, masses are very close to
each other leading to enhanced coannihilation through the processes
$\Neu 1 \Cha 1 \to q_u\, \wbar q_d$ and $\Neu 1\Neu 1, \Cha 1 \Cha 1
\to q\, \wbar q, \ell \wbar \ell, W^+ W^-$, which allows the
acceptable parameter space to extend to very large values of \M2 and
\m0.

The \WMAP-preferred region is very wide as compared to, \eg, the universal
model or other models in SU(5) \cite{Huitu:2008sa,Huitu:2007vw}. The
spectrum is relatively heavy for the \WMAP-preferred region, around a
couple of TeV.  This leads to a wide range in the parameter space, but
from the point of view of the coming Large Hadron Collider such a mass
spectrum may be problematic.
However, the representation {\bf 210} produces naturally a neutralino
with mass around a TeV, which seems to be favorable in the view of
the recent PAMELA \cite{Adriani:2008zq} and ATIC \cite{:2008zz}
results of the excess positron and positron+electron flux. Under
certain circumstances, a nearby clump of 600-1000 GeV neutralino LSP
could fit into these observations \cite{Hooper:2008kv}.

\begin{figure}
    \includegraphics[width=0.32\textwidth]{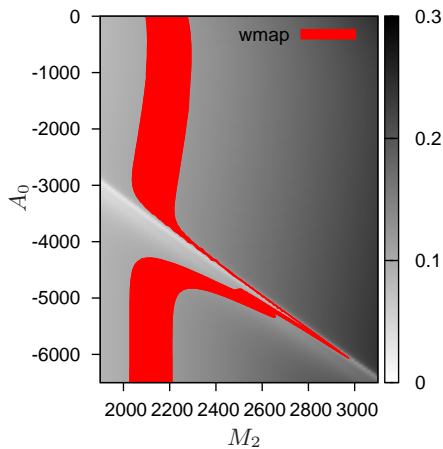}
  \caption{Relic density in \textbf{Rep 210}: $\tan\beta = 10,\ \mathrm{sgn} ( \mu ) =
    +1,\ m_0=1$ TeV. The collider constraints are fulfilled.}
  \label{fig:relic210A0}
\end{figure}
In Fig.~\ref{fig:relic210A0} the area of preferred thermal relic
density in the representation \textbf{210} is plotted for the same set
of (GUT scale) parameters as in Fig.~\ref{fig:relic210m0}, except that
now the trilinear $A_0$-parameter is varied along the y-axis, and the
$m_0$ is set to 1 TeV. The interesting feature in this figure is the
existence of the pseudoscalar Higgs annihilation channel through the
$\M2$-values. This reduces greatly the relic density in the parameter
space where the LSP mass equals or is less than half of the $A$-Higgs
mass. This has the effect of pushing the \WMAP-preferred relic density
region to heavier neutralino masses, and therefore to larger \M2
values. The width of the \WMAP region is naturally the same as before,
since the top of the figure, where $A_0=0$, coincides with 
Fig.~\ref{fig:relic210m0} with $m_0=1000$ GeV. With these parameters,
the NLSP is always the lighter chargino, and the lightest neutralino
is dominantly a wino, which can also be read from
Fig.~\ref{fig:relic210-c}.

\section{Discussion and summary}
We studied the dark matter allowed regions in the SO(10) GUT
representations, of which all but the singlet may lead to
nonuniversal gaugino masses.  The \WMAP-preferred relic density
regions are quite distinct for different representations, thus leading
to quite different particle spectra for each representation.

In the representation {\bf 54}, the lightest neutralino is
predominantly a bino, leading to the narrow areas of the
\WMAP-favored region. The excessive relic density is diluted either
by increasing the Higgsino component or by coannihilation with other
particles. The breaking chain \SUSO7 predicts an upper limit for the
lighter chargino mass for the chosen parameters. For part of the \WMAP
allowed region, the whole neutralino and chargino spectrum is lighter 
than the spectrum of sfermions.
For the \SUPS breaking chain, the relic density area is narrow in the
parameter space.  Interestingly, there may exist a region, where the
stau, smuon and the lightest neutralino masses are in a very close
range to each other. This can lead to long-lived staus and smuons,
which may be stable in the collider time scale.

In the {\bf 210} dimensional representation the lightest neutralino is
either wino or Higgsino, which leads to a low thermal relic
density. In addition, the lightest chargino and the lightest
neutralino tend to be close in mass, thus providing a coannihilation
channel. The preferred relic density area is quite large.  The
sparticle spectrum is heavy, as compared to the universal mSUGRA case.
Only in a small part of the \WMAP-preferred parameter region are a few \SUSY
particles expected to be within the kinematic reach of the \LHC.

In this work we have studied each representation separately.  It is
obvious that if several representations affect simultaneously the
composition of neutralinos, the possible \WMAP-preferred region in the
parameter space may be relaxed.  However, if the neutralinos and
charginos are found with a certain mass pattern, it helps to
understand the relation of the lightest neutralino with dark matter,
if characteristics of each representation are known.

\section{Acknowledgments}
The work of K.H.~is supported by the Academy of Finland (Project
No.~115032). The work of J.L.~is supported by the Bundesministerium
f\"ur Bildung und Forschung, Berlin-Bonn.

\bibliography{bib-relicdensity-so10_v1.5}

\end{document}